\documentstyle[12pt]{article} 
\begin{document}
\begin{titlepage}
\rightline{\vbox{\halign{&#\hfil\cr
&CERN-TH 6785/93 \cr
&ANL-HEP-PR-93-7\cr
&January, 1993\cr}}}
\vspace{0.5in}
\begin{center}

{\Large\bf
Longitudinal Structure Function\\ at Intermediate x and the Gluon Density
}
\medskip
\medskip
\medskip
\medskip

\normalsize Edmond L. Berger$^{1,2}$ and Ruibin Meng$^2$
\\ \smallskip
\medskip
{$^1$CERN-Geneva\\}
\medskip
$^2$High Energy Physics Division\footnote{Work supported by the U.S.
Department of Energy, Division of High Energy Physics,
Contract\newline W-31-109-ENG-38.} \\
Argonne National Laboratory\\Argonne, IL 60439\\
\medskip
\end{center}

\vspace{1.0in}

\begin{abstract}
Calculations are presented of the longitudinal structure function
$F_L(x, Q^2)$.
We use next-to-leading order expressions in QCD $({\cal{O}}(\alpha_s^2))$
plus parton densities determined previously from global fits to data on deep
inelastic lepton scattering, prompt photon production, and bottom quark
production.  Anticipated data from the DESY ep collider HERA should
provide discriminating information on the gluon density,
particularly for values of $x < 0.05$.
\end{abstract}
\end{titlepage}

\newpage

\indent In perturbative quantum chromodynamics(QCD), the longitudinal structure
function $F_L(x, Q^2)$, measured in deep inelastic lepton scattering, may be
expressed to leading order as a sum of integrals over the gluon and quark
densities of the nucleon[1].  As usual $Q^2$ is the square of the four-vector
momentum exchange, and $x$ is the Bjorken scaling variable.  At small values
of $x$, the gluon contribution dominates, and, to fair approximation,

\begin{eqnarray}
F_L (x,Q^2) \cong \frac{2 \alpha_s(Q^2)}{3 \pi} \frac{1}{1.77}
\frac{x}{a} G(\frac{x}{a},Q^2).
\end{eqnarray}

\noindent
Here $G(x, Q^2)$ is the gluon density, $\alpha_s(Q^2)$ is the QCD
coupling strength, and $a$ is a parameter whose value is
about $0.4$ for $F_L$[2].  Correspondingly, measurement of $F_L$ has long been
advocated as perhaps the most direct probe of the gluon density at small
$x$.  Studies indicate that precise measurements of $F_L$ should be feasible
at the HERA ep collider at DESY for values of $x$ of order $10^{-2}$, or less,
and values of $Q^2$ in the range $10$ to $100~$GeV$^2$[3].

Recently a full next-to-leading order calculation
$({\cal{O}}(\alpha_s^2))$
in QCD has been published for the deep inelastic structure functions
$F_2$ and $F_L$[4].
These analytic expressions allow us to extend the exploration of the
sensitivity  of $F_L$ to G beyond lowest order. In addition, new sets of parton
densities evolved at two-loop level[5,6]
have been determined from global fits to recent data from deep
inelastic muon and neutrino scattering[7], prompt photon production[8], and,
for the first time[5] bottom quark production cross sections at collider
energies[9].  There are significant and interesting differences among the gluon
densities in these different fits.  We use these various distributions in our
calculations of $F_L$, and, as we will show, measurements of $F_L$ at HERA
should be discriminating and instructive at both intermediate $(0.005 <
x < 0.1)$ and smaller $x$.

Complementary to prior phenomenological studies [10], we call attention in this
paper principally to the region of intermediate $x$.  As we have described
previously[5], within the context of ${\cal O}(\alpha_s^3)$ perturbative
QCD, an acceptable fit to the CDF data on bottom quark production at
Fermilab hadron collider energies requires that the gluon density be enhanced
in the vicinity of $x = 0.05$, relative to the value it tends to take on if
these data are omitted from the fit.  No other data yet constrain the gluon
density  directly in the neighborhood of $x = 0.05$.
As we will show, the enhancement in
$G(x, Q^2)$ shows up clearly in $F_L(x, Q^2)$.
Distinct from an enhanced gluon
density at intermediate $x$, alternative theoretical contributions, including
the role of off-shell initial gluons [11], may help explain the bottom quark
data.  Thus, independent information on $G(x, Q^2)$ from data on $F_L$ is
highly desirable.

In all global fits done to extract parton densities, specific functional forms
are adopted for the gluon density $G(x, Q_0^2)$ and other parton densities
at the reference value $Q_0^2$.  Free parameters are varied to obtain a
best fit to the  data set selected.
(The chosen functional forms are somewhat limiting since
they may not in all cases be versatile enough to match the form of the data).
One constraint imposed on the parton densities is the momentum sum rule.  If
attention is restricted to the gluon density, this sum rule takes the
approximate form

\begin{eqnarray}
\int_0^1 x G(x, Q^2) dx \cong 0.5
\end{eqnarray}

\noindent
Correspondingly, an enhancement of $G(x, Q^2)$ in the neighborhood of $x
= 0.05$ necessarily requires diminution of $G(x, Q^2)$ elsewhere in
$x$, either at large $x$ or at small $x$.  In all of our global fits[5],
we require good agreement
with data on deep inelastic lepton scattering, including the most recent
muon and neutrino data from the NMC and CCFR collaborations [7].
Experimental support for the gluon density at
$x > 0.2$ is provided by data on the production of prompt photons in
nucleon-nucleon interactions at fixed target
energies[8].  We have determined two different sets of parton densities[5],
sets A and B,
that differ according to the extent to which we insist on a good fit to the
prompt photon data.  In Tables 1 and 2, we provide the values of chi-squared
for these
two fits, and in Fig. 1 we compare the gluon densities from the two fits at
$Q = 5~$GeV.  Although hadronic jet data were not included in the fits,
it is of interest that the two sets of densities yield
good fits[12] to the inclusive hadronic
jet cross section data at $\sqrt{s} = 1.8~$TeV[13] and do as well as other
published densities at representing the jet data at $\sqrt{s} =
546~$GeV[14].

The prompt photon data require a relatively large gluon density for $x >
0.2$. As indicated in Fig. 1, set A that provides the best simultaneous
representation of the bottom quark and prompt photon data has a gluon density
that is relatively suppressed at small $x$.  The relative
suppression is most apparent at small values of $Q^2$; differences tend to be
washed out after evolution to large $Q^2$, as is apparent in the curves of
$F_L(x, Q^2)$ discussed below.

The analytic behavior of parton densities in the region of very small $x$ has
been the subject of considerable theoretical discussion.  As emphasized above,
our interest is focused here principally on the region of intermediate $x$.
Nevertheless, it is useful to summarize the nature of the debate at very small
$x$.  The ``conventional'' behavior $x G (x,Q_0^2) \rightarrow {\rm constant}$
has been assumed in many parametrizations over the years,
motivated by considerations founded in
Regge pole phenomenology.  More recently, the ``singular''  behavior $x G
(x,Q_0^2) \rightarrow x^{-1/2}$  has been suggested on the basis of
studies[15] of the Lipatov
evolution equation, and sets of parton densities incorporating this analytic
form have been derived.  We have not made fits in which the singular form was
imposed, and, as a result, we have little to add to earlier work [10] showing
that data on $F_L (x,Q^2)$ below $x = 0.005$ would help to distinguish the
conventional and singular behaviors.  The relative enhancement to which we call
attention should begin to be apparent at more easily accessible intermediate
values of $x$, $0.005 < x < 0.1$.

The complete expressions for $F_L$ in order $\alpha_s^2$ may be found in the
paper of Zijlstra and van Neerven [4].  Owing to their length we do not
reproduce them here.  The analysis in this paper is carried out in the
$\overline{\rm MS}$ factorization scheme.
We assume five flavors of initial quarks with
mass corrections for the bottom quark.

Our numerical results are presented in Fig. 2 as functions of $x$ for four
values of $Q^2$: 10, 40, 80, and 160~GeV$^2$.
We show predictions for $F_L(x, Q^2)$
obtained from our sets A and B along with expectations from the non-singular
set $D_0^\prime$ of MRS.
The enhancement in the vicinity of $x= 0.05$ in the gluon densities of sets A
and B, visible in Fig. 1, is also apparent in the curves of $F_L (x,Q^2)$
for all four values of $Q^2$.  These results strengthen and extend to a larger
domain in $x$ previous conclusions [3,10] that measurements of $F_L$ will
provide discriminating information on $G(x, Q^2)$.
(We caution that owing to the
important role of the ${\cal O}(\alpha_s^2)$ contributions, the leading order
approximation in Eq. (1) should not be applied to deduce $G(x, Q^2)$ from the
data.)

As remarked above, the enhancement in the gluon density near $x = 0.05$
arises from ${\cal O}(\alpha_s^3)$ QCD fits to data on bottom quark production
at $\sqrt{s} = 1.8~$TeV.
Alternative explanations of the bottom quark data may involve contributions
to the cross section in
order $\alpha_s^4$ and beyond, including the effects of off-shell initial
gluons[11].
An independent determination of $G(x, Q^2)$ from data on $F_L(x, Q^2)$
would therefore be particularly instructive.  Knowledge of the gluon density
at small $x$ is of further value for more reliable estimates of rates for hard
scattering processes at supercollider energies[16].

We are grateful to Professor W.L. van Neerven for discussions and for providing
a copy of his code for computing the ${\cal O}(\alpha_s^2)$ contributions
to $F_L$.

\newpage

\section*{References}

\begin{enumerate}

\item\label{1}C.~G.~Callan and D.~J.~Gross, Phys. Rev. Lett. {\bf 22},
156 (1969); G.~Altarelli and G.~Martinelli, Phys. Lett. B{\bf 76},
89 (1978).
\item\label{2} J.~Bl\"umlein in {\it Proceedings of the Large Hadron
Collider Workshop, Achen, October 4-9, 1990}, eds. G.~Jarlskog and
D.~Rein, vol. 2, p.850; N.~Magnussen and G.~A.~Schuler in {\it Proceedings
of the Large Hadron Collider Workshop, October 4-9, 1990}, eds. G.~Jarlskog
and D.~Rein, vol. 2, p.858.
\item\label{3} M.~Cooper-Sarkar {\it et al.}, Z. Phys.
C{\bf 39}, 281 (1988).
\item\label{4} E. B. Zijlstra and W. L. van Neerven, Nucl. Phys. {\bf
B383}, 525 (1992).
\item\label{5} E.~L.~Berger, R.~Meng, and J.-W.~Qiu,
ANL-HEP-CP-92-79, to be published in the Proc. XXVI Int. Conf. on High Energy
Physics, Dallas, 1992.   E.~L.~Berger and R.~Meng, ANL-HEP-CP-92-108, to
be published in the Proc. DPF92, Fermilab, 1992.
\item\label{6} A.D.Martin, W.J.Stirling, and R.G.Roberts,
Durham/Rutherford report
RAL-92-078, DTP/92/80, November 1992.
\item\label{7}BCDMS Collab., A. C. Benvenuti {\it et al.}, Phys.
Lett.  {\bf B223}, 485 (1989); {\bf B237}, 599 (1990). CDHSW Collab.,
J.~P.~Berge {\it et al.}, Zeit. Phys. C{\bf 49}, 187 (1990).
NMC Collab., P. Amaudruz {\it et al.}, CERN-PPE/92-124.
CCFR Collab., S. R. Mishra {\it et al.}, Nevis Preprint 1459.
\item\label{8} WA70 Collab., M. Bonesini {\it et al.}, Zeit. Phys. C{\bf 37},
535 (1988); C{\bf 38}, 371 (1988).
E706  Collab., G. Alverson {\it et al.}, Phys. Rev. Lett.
{\bf 68}, 2584 (1992).
\item\label{9} CDF Collaboration, F.~Abe {\it et al.},
Phys.~Rev.~Lett.~{\bf 68}, 3403 (1992);
Phys.~Rev.~Lett.~{\bf 69}, 3704 (1992).
UA1 Collab., C. Albajar
{\hbox{\it et al.}}, Phys. Lett. {\bf B256}, 121 (1991).
\item\label{10} e.g. L. Orr and W.J.Stirling, Phys. Rev. Lett. {\bf 66},
1673 (1991).
\item\label{11} J. C. Collins and R. K. Ellis, Nucl. Phys. {\bf B360}, 3
(1991);
S. Catani {\it et al.}, Nucl. Phys. {\bf B366}, 135 (1991);
E. M. Levin {\it et al.}, 
Sov. J. Nucl. Phys. {\bf 54}, 867 (1991).
\item\label{12}S.~D.~Ellis and D.~E.~Soper, private communication;
S.~D.~Ellis, Z.~Kunszt and D.~E.~Soper
Phys.~Rev.~Lett.~{\bf 64}, 2121 (1990);
S.~D.~Ellis, Z.~Kunszt and D.~E.~Soper
Phys.~Rev.~Lett.~{\bf 69}, 3615 (1992).
\item\label{13}CDF Collaboration, F.~Abe {\it et al.},
Phys.~Rev.~Lett.~{\bf 68}, 1104 (1992).
\item\label{14}CDF Collaboration, F.~Abe {\it et al.},
Fermilab-PUB-92-286-E, submitted to Phys.~Rev.~Lett.
\item\label{15} J. Collins, {\it Proceedings of the Summer Study on the Design
and Utilization of the SSC, Snowmass, 1984}, edited by R.~Donaldson and
J.~Morfin, pp. 251--255;  L.N. Lipatov in
{\it Perturbative QCD}, edited by A.~H.~Mueller (World Scientific, Singapore,
1989), p. 411.
\item\label{16}E.~L.~Berger and R.~Meng, Phys. Rev. D{\bf46}, 169 (1992).
\end{enumerate}

\newpage

\begin{table*}[t]
\caption{Values of $\chi^2$ from the combined fit A to
the prompt photon, bottom quark, deep inelastic
lepton scattering
and massive lepton pair production data.
The fit to the deep inelastic data is
restricted to points with $Q > 3.16~$GeV and $W > 4~$GeV.
The top line specifies the
data sets,
and the second line lists the values of $\chi^2$ divided by the number of data
points.
The BCDMS data are renormalized by a factor 0.985.
The NMC data are renormalized by a factor 1.020.
The CCFR data
are renormalized by a factor 0.965.
}
\begin{center}
\vspace{0.1in}
Set A
\begin{tabular}{lccccccccc}
\noalign{\vspace{12pt}}
\hline\hline
& UA1 & CDF & WA70 & E706 & BCDMS & CDHSW
& NMC & CCFR & E605\\
&     &     &      &      & $F^H_2$\ \ \ \ $F^D_2$ & $F_2$\ \ \ \ $xF_3$
& $F^H_2$\ \ \ $F^D_2$ & $F_2$\ \ $xF_3$ &  \\
\noalign{\vspace{2pt}}
\hline
\noalign{\vspace{2pt}}
&0.69 &1.08 &1.28 & 1.45 & 0.65\ \ \ 1.05\ \ & 0.47\ \ 0.52\
& 1.52\ \ 1.35 & \ \ \ 1.85\ \ \ \ \ 0.76 & 1.06\\
 \hline\hline
\end{tabular}
\end{center}
\end{table*}

\begin{table*}[t]
\caption{Values of $\chi^2$ from the combined fit B to
the bottom quark, deep inelastic
lepton scattering
and massive lepton pair production data.
The fit to the deep inelastic data is
restricted to points with $Q > 3.16~$GeV and $W > 4~$GeV.
The top line specifies the
data sets,
and the second line lists the values of $\chi^2$ divided by the number of data
points.
The BCDMS data are renormalized by a factor 0.985.
The NMC data are renormalized by a factor 1.020.
The CCFR data
are renormalized by a factor 0.965.
}
\begin{center}
\vspace{0.1in}
Set B
\end{center}
\begin{center}
\begin{tabular}{lccccccc}
\hline\hline
& UA1 & CDF & BCDMS & CDHSW
& NMC & CCFR & E605\\
&     &      & $F^H_2$\ \ \ \ $F^D_2$ & $F_2$\ \ \ \ $xF_3$
& $F^H_2$\ \ \ $F^D_2$ & $F_2$\ \ $xF_3$ &  \\
\noalign{\vspace{2pt}}
\hline
\noalign{\vspace{2pt}}
&1.02 &0.86 & 0.67\ \ \ 1.02\ \ & 0.50\ \ 0.53\
& 1.61\ \ 1.41 & \ \ \ 1.86\ \ \ \ \ 0.77 & 1.14\\
 \hline\hline
\end{tabular}
\end{center}
\end{table*}

\vfill\eject

\newpage

\samepage{
\section*{Figure Captions}
\smallskip
\begin{itemize}
\item[{Fig. 1.}]
Solid and dashed curves show the behavior of the gluon densities
from our fits A and B as a function of $x$ for $Q = 5~$GeV.  They are compared
to the density (dotted) from the conventional set $D_0^\prime$ of MRS, ref.
[6].
\item[{Fig. 2.}]
Solid and dashed curves show the predicted behavior of the
longitudinal structure function $F_L(x, Q^2)$ derived
from our fits A and B as a function of $x$ at the $Q^2$ values a)
$10~$GeV$^2$, b)
$40~$GeV$^2$, c) $80~$GeV$^2$, and d) $160~$GeV$^2$.
Also shown are values calculated
from the conventional set $D_0^\prime$ of MRS, ref. [6].
\end{itemize}
}
\end{document}